\documentstyle[aps,prl,floats,graphicx,times]{revtex}

\begin{document}
\draft
\wideabs{
\title{Pairing Fluctuation Theory of Superconducting Properties in 
Underdoped to Overdoped Cuprates}
\author{ Qijin Chen, Ioan Kosztin, Boldizs\'ar Jank\'o, and K. Levin}
\address{The James Franck Institute, The University of Chicago, 5640
  South Ellis Avenue, Chicago, Illinois 60637}
\date{\today}
\maketitle

\begin{abstract}
  We propose a theoretical description of the superconducting state of
under- to overdoped cuprates, based on the short coherence length of
these materials and the associated strong pairing fluctuations.
  The calculated $T_c$ and the zero temperature excitation gap
  $\Delta(0)$, as a function of hole concentration $x$, are in
  semi-quantitative agreement with experiment.
  Although the ratio $T_c/\Delta(0)$ has a strong $x$ dependence,
  different from the universal BCS value, and $\Delta(T)$ deviates
  significantly from the BCS prediction, we obtain, quite remarkably,
  quasi-universal behavior, for the normalized superfluid density
  $\rho_s(T)/\rho_s(0)$ and the Josephson critical current
  $I_c(T)/I_c(0)$, as a function of $T/T_c$.
  While experiments on $\rho_s(T)$ are consistent with these results,
future measurements on $I_c(T)$ are needed to test this prediction.
\end{abstract} 

\pacs{PACS numbers: 
74.20.-z,
74.25.-q,
74.62.-c,
74.72.-h
\hfill
\textbf{cond-mat/9807414}
}
}
Pseudogap phenomena in the cuprates are of interest not only because of
the associated unusual normal-state properties, but more importantly
because of the constraints which these phenomena impose on the nature of
the superconductivity and its associated high $T_c$. Moreover, this
superconducting state presents an interesting challenge to theory: while
the normal state is highly unconventional, the superconducting phase
exhibits some features of traditional BCS superconductivity along with
others which are strikingly different.
 
Thus far, there is no consensus on a theory of cuprate
superconductivity. Scenarios which address the pseudogap state below
$T_c$ can be distinguished by the character of the excitations
responsible for destroying superconductivity.  In the theory of Lee and
Wen \cite{LeeWen}, the destruction of the superconducting phase is
associated with the excitation of the low-lying
quasiparticles near the $d$-wave gap nodes.  By contrast, Emery and
Kivelson \cite{Emery} argue that the destruction of the
superconductivity is associated with low frequency, long wavelength
phase fluctuations within a microscopically inhomogeneous model, based
on one dimensional ``stripes".

In the present paper, we present an alternative scenario in which, along
with the quasiparticles of the usual BCS theory, there are
additionally incoherent (but \textit{not pre-formed}) pair excitations
of finite momentum \textbf{q}, which assist in the destruction of
superconductivity.  This approach is based on a self-consistent
treatment of the coupling of single particle and pair states. It
represents a natural extension of BCS theory to the short coherence
length ($\xi$) regime and provides a quantitative framework for
addressing cuprate superconductivity.  Here, we find a pronounced
departure from BCS behavior in the underdoped limit which is
continuously reduced with increasing hole concentration $x$.  We derive
a phase diagram for $T_c$ and the zero temperature gap, $\Delta(0)$, as
a function of $x$, which is in semi-quantitative agreement with (the
anomalous) behavior observed in cuprate experiments, and we compute
properties of the associated superconducting state such as the
superfluid density $\rho_s$ and Josephson critical current $I_c$.  When
these are plotted as $\rho_s(T)/\rho_s(0)$ and $I_c(T)/I_c(0)$, as a
function of $T/T_c$, we deduce a quite remarkable, nearly universal
behavior for the entire range of $x$.
 
As a simple model for the cuprate band structure, we consider a
tight-binding, anisotropic dispersion $\epsilon_{\mathbf{k}} =
2\,t_\parallel (2-\cos{k_x}-\cos{k_y}) + 2\;t_{\perp}(1-\cos{k_{\perp}})
- \mu\,$, where $t_\parallel$ ($t_{\perp}$) is the hopping integral for
the in-plane (out-of-plane) motion and $t_\perp \ll t_\parallel$
\cite{c-axis}.
We assume that the fermions interact via an effective pairing
interaction with $d$-wave symmetry $V_{\mathbf{k,k'}} =
-|g|\varphi_{\mathbf{k}}\varphi_{\mathbf{k'}}$ so that $
\varphi_{\mathbf{k}} = \frac{1}{2}\left(\cos{k_x}-\cos{k_y}\right)$.
The present approach is built on previous
work\cite{Janko-Maly,Qijin,Kosztin} based on a particular diagrammatic
theory, first introduced by Kadanoff and Martin \cite{kadanoff61}, and
subsequently extended by Patton\cite{Patton}.  This approach can be used
to describe the widely discussed BCS to Bose-Einstein cross-over
problem\cite{Crossover}, which has been associated with small $\xi$. The
``pairing approximation'' of Refs.~\onlinecite{kadanoff61,Patton} leads
to 

\begin{mathletters}
  \label{eq:pa}
  \begin{eqnarray}
    \label{eq:pa1}
    \Sigma(K) &=&  \sum_Q t(Q)\,G_o(Q-K)\,\varphi^2_{{\mathbf{k}}-
{\mathbf{q}}/2}\;,\\ 
    \label{eq:pa2}
    g &=& [1+g\chi(Q)]t(Q)\;,
  \end{eqnarray}
\end{mathletters}
where $\Sigma(K)$ is the self-energy, $\chi(Q)=\sum_K G(K)\, G_o(Q-K)
\varphi^2_{{\mathbf{k}}-{\mathbf{q}}/2}$ is the pair susceptibility.
Equations (\ref{eq:pa}), along with the number equation $n = 2\sum_K
G(K)$, self-consistently determine both the Green's function $G(K)$ and
the pair propagator, i.e., T-matrix $t(Q)$.
We use a four-vector notation, e.g., $K\equiv(\bbox{k};i\omega)$,
$\sum_K \equiv T\sum_{i\omega,\mathbf{k}}$ and $G_o(K)=(i\,\omega -
\epsilon _{\mathbf{k}})^{-1}$.
We now show that these equations yield a natural extension of BCS
theory to include incoherent pairs (labeled by ${pg}$), along with the
usual quasiparticles and superconducting condensate
(labeled by ${sc}$).

We write the T-matrix and self-energy below $T_c$ as
$t(Q) = t_{sc}(Q)+t_{pg}(Q),\, \text{and}\;  \Sigma(K) =
\Sigma_{sc}(K) + \Sigma_{pg}(K)\,.$
The condensate contribution assumes the familiar BCS form
$t_{sc}(Q) = -\delta(Q)\Delta_{sc}^2/T$, where $\Delta_{sc}$ is the
superconducting gap parameter (and can be chosen to be real) and
$\Sigma_{sc}(K) = \Delta_{sc}^2\, \varphi_{\mathbf{k}}^2/\left(i\omega +
\epsilon_{\mathbf{k}}\right)$.  Inserting the above forms for the
T-matrix into Eq.~(\ref{eq:pa2}), one obtains the gap equation
$1+g\chi(0)=0$, as well as (for any non-zero $Q$), $t_{pg}(Q) =
g/\left(1+g\chi(Q)\right)$.
Note that because of the gap equation, $t_{pg}(Q)$ is highly peaked
about the origin, with a divergence at $Q=0$ \cite{thouless}. As a
consequence, in evaluating the associated contribution to the
self-energy, the main contribution to the $Q$ sum comes from this
small $Q$ divergent region so that $\Sigma_{pg}(K) \approx -G_o(-
K)\Delta_{pg}^2\varphi^2_{\mathbf{k}}$, where we have defined the
pseudogap parameter\cite{PG_fluct}

\begin{mathletters}
  \label{eq:gap}
\begin{equation}
  \label{eq:gap3}
 \Delta^2_{pg} \equiv -\sum_Q t_{pg}(Q) =
 -\sum_Q\frac{g}{1+g\chi(Q)}  \;.  
\end{equation}
Thus, both $\Sigma_{pg}$ and the total self-energy $\Sigma$ can be well
approximated by a BCS-like form, i.e., $\Sigma(K) \approx
\Delta^2\,\varphi^2_{\mathbf{k}}/\left(i\omega_n+
\epsilon_{\mathbf{k}}\right)$, where
$\Delta\equiv\sqrt{\Delta_{sc}^2+\Delta_{pg}^2}$ is the magnitude of the
total excitation gap, with the \textbf{k} dependence given by the
$d$-wave function $\varphi_{\mathbf{k}}$. Within the above
approximations, the gap and number equations reduce to
\begin{eqnarray}
  \label{eq:gap1} 1+g\sum_{\mathbf{k}} \frac{1-
    2f(E_{\mathbf{k}})}{2E_{\mathbf{k}}}\, \varphi^2_{\mathbf{k}} &=& 0\;,
  \\ \label{eq:gap2} \sum_{\mathbf{k}}\left[1-
    \frac{\epsilon_{\mathbf{k}}}{E_{\mathbf{k}}} +
    \frac{2\epsilon_{\mathbf{k}}}{E_{\mathbf{k}}}\,f(E_{\mathbf{k}})\right]
    &=& n\;,
\end{eqnarray}
\end{mathletters}
where the quasiparticle energy dispersion
$E_{\mathbf{k}}=(\epsilon_{\mathbf{k}}^2 +
\Delta^2\varphi^2_{\mathbf{k}})^{1/2}$ contains the full excitation gap
$\Delta$.

The set of equations (\ref{eq:gap}) can be used to determine the
superconducting transition temperature $T_c$ (where $\Delta_{sc} = 0$),
and the temperature dependence of the various gap parameters.
Eq.~(\ref{eq:gap3}) contains the physics of the pair excitations, or
pseudogap. The remaining two Eqs.~(\ref{eq:gap1}) -
(\ref{eq:gap2}) are analogous to their BCS counterparts but with a
finite (as a result of non-zero $\Delta_{pg}$) excitation gap at $T_c$.

It should be stressed that physical quantities which characterize the
superconducting state depend on the pair and particle excitations, as
well as condensate in different ways. The superfluid density can be
written in terms of the London penetration depth as
$\rho_{s,ab}(T)/\rho_{s,ab}(0) =
\left[\lambda_{ab}(0)/\lambda_{ab}(T)\right]^2$, 
where 

\begin{eqnarray}
  \label{rhos}
  \lambda_{ab}^{-2} &=& \frac{4\pi 
e^2\Delta_{sc}^2}{c^2}\sum_{\mathbf{k}}
  \frac{\varphi_{\mathbf{k}}}{E_{\mathbf{k}}^2}\left[
    \frac{1-2f(E_{\mathbf{k}})}{2E_{\mathbf{k}}} + f^\prime
    (E_{\mathbf{k}})\right] \\
  && \times\left[ \varphi_{\mathbf{k}} \left( \frac{\partial
        \epsilon_{\mathbf{k}}}{\partial {\mathbf{k}}_\parallel}\right)^2 
-
    \epsilon_{\mathbf{k}} \frac{\partial \epsilon_{\mathbf{k}}}
    {\partial{\mathbf{k}}}\cdot \frac{\partial
      \varphi_{\mathbf{k}}}{\partial{\mathbf{k}}} \right]\;.  
\nonumber
\end{eqnarray}
During the calculation special attention should be paid to lattice
effects\cite{Misawa} and to the vertex correction (associated with the
pseudogap self-energy) which enforces gauge invariance via the
generalized Ward identity.  This identity insures that $\rho_s \propto
\Delta_{sc}^2$ and it vanishes identically at and above $T_c$.
The prefactor $\Delta_{sc}^2 = \Delta^2 - \Delta_{pg}^2$ in
Eq.~(\ref{rhos}) indicates that pairs (in addition to
quasiparticles) serve to reduce the superfluid density.

In a related fashion, we address c-axis Josephson tunneling between two
identical high $T_c$ superconductors. This situation is relevant to both
break junction experiments\cite{JohnZ} and to intrinsic Josephson
tunneling\cite{IJJ} as well.  An expression for the Josephson critical
current\cite{AB63} can be derived under the presumption that the
tunneling matrix element $|T_{\mathbf{kp}}|^2= \left| T_0 \right|^2
\delta_{{\mathbf{k}}_\parallel{\mathbf{p}}_\parallel}+ \left| T_1
\right|^2$, where only the first (coherent) term contributes for a
$d$-wave order parameter,

\begin{eqnarray}
  \label{crit}
  I_c &=& 2e|T_0|^2 \Delta_{sc}^2 \sum_{\mathbf{kp}}
  \delta_{{\mathbf{k}}_\parallel{\mathbf{p}}_\parallel}
  \frac{\varphi_{\mathbf{k}} \varphi_{\mathbf{p}}}
  {E_{\mathbf{k}}E_{\mathbf{p}}} \\
  && \times \left[ \frac{1-f(E_{\mathbf{k}})-f(E_{\mathbf{p}})}
    {E_{\mathbf{k}}+E_{\mathbf{p}}}+
    \frac{f(E_{\mathbf{k}})-f(E_{\mathbf{p}})} {E_{\mathbf{k}}-E_{\mathbf{p}}}
  \right]\;.\ \nonumber
\end{eqnarray}
Equation~(\ref{crit}), like Eq.~(\ref{rhos}), differs from the usual
BCS form (as well as that assumed by Lee and
Wen\cite{LeeWen,Millis}) in that the prefactor $\Delta_{sc}^2$ is no
longer the total excitation gap $\Delta^2$.

The remainder of this paper is directed towards understanding three
experimental characteristics of the cuprates: (i) the phase diagram,
(ii) the superfluid density and (iii) the Josephson critical current.

(i) In order to generate physically realistic values of the various
energy scales, we make two assumptions: (1) We take $g$ as
doping-independent (which is not unreasonable in the absence of any more
detailed information about the pairing mechanism) and (2) incorporate 
the effect of the Mott transition at half filling, by introducing
an $x$-dependence into the in-plane hopping matrix elements
$t_\parallel$, as would be expected in the limit of strong on-site
Coulomb interactions in a Hubbard model\cite{Anderson87}. Thus the
hopping matrix element is renormalized as $t_\parallel(x) \approx t_0
(1-n)=t_0x$, where $t_0$ is the matrix element in the absence of Coulomb
effects.  This $x$ dependent energy scale is consistent with the
requirement that the plasma frequency vanish at $x = 0$. These
assumptions leave us with one free parameter $-g/4t_0$, for which we
assign the value 0.15 to optimize the overall fit of the phase diagram
to experiment. We take $t_\perp/t_\parallel \approx 0.01$
\cite{Anisotropy}, and $t_0 \approx 0.6$~eV, which is reasonably
consistent with experimentally based estimates\cite{t0}.

\begin{figure}
\centerline{\includegraphics{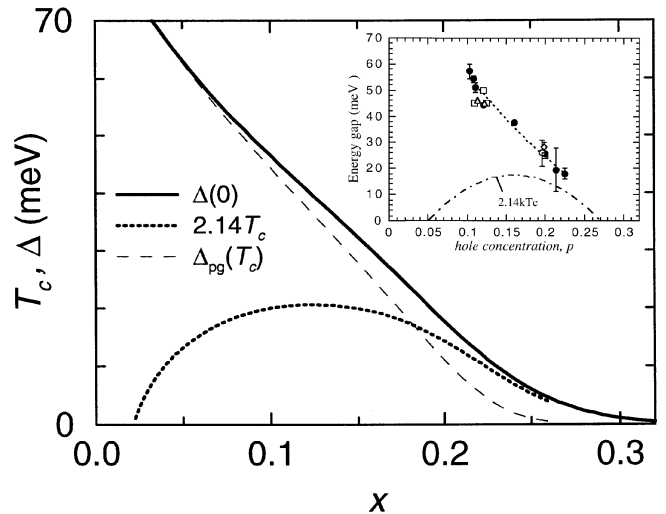}} 
\medskip
\caption{Phase diagram showing $\Delta (0)$ and $T_c$ as well as
  $\Delta_{pg} (T_c)$. The inset shows experimental results 
  from Ref.~\protect\onlinecite{JohnZ}.}
\label{Fig1}
\end{figure}

The results for $T_c$, obtained from Eqs.~(\ref{eq:gap}), as a function
of $x$ are plotted in Fig.~\ref{Fig1}. Also indicated is the
corresponding zero temperature excitation gap $\Delta (0)$ as well as
the pseudogap $\Delta_{pg}$ at $T_c$.  These three quantities provide
us, for use in subsequent calculations, with energy scales which are in
reasonable agreement with the data of Ref.~\cite{JohnZ}, shown in the
inset.  The temperature dependences of the energy gaps in Fig.~1 are
shown as the lower inset to Fig.~2, for a slightly underdoped case with
$x=0.125$.  The relative size of $\Delta_{pg} (T_c) $, compared to
$\Delta (0)$, increases with decreasing $x$.  In the highly overdoped
limit this ratio approaches zero, and the BCS limit is recovered. This
inset illustrates the general behavior as a function of $T/T_c$: the
excitation gap $\Delta$ is, generally, finite at $T_c$ , the
superconducting gap $\Delta_{sc}$ is established at and below $T_c$,
while the pseudogap $\Delta_{pg}$ decreases to zero as $T$ is reduced
from $T_c$ to 0. This last observation is consistent with general
expectations for $\Delta^{2}_{pg} \approx \left\langle
|\Delta|^2\right\rangle - \left|\Delta_{sc}\right|^2$ \cite{PG_fluct}.

It is important to stress, that our subsequent results for the
superfluid density and Josephson current, need not be viewed as
contingent on the detailed $x$-dependence used to derive the phase
diagram.  One can approach the calculations of these quantities by
taking $T_c(x)$ and the various gap parameters (shown in the inset) as
phenomenological inputs, within the context of the present formalism.

\begin{figure}
\centerline{\includegraphics[width=3.4in]{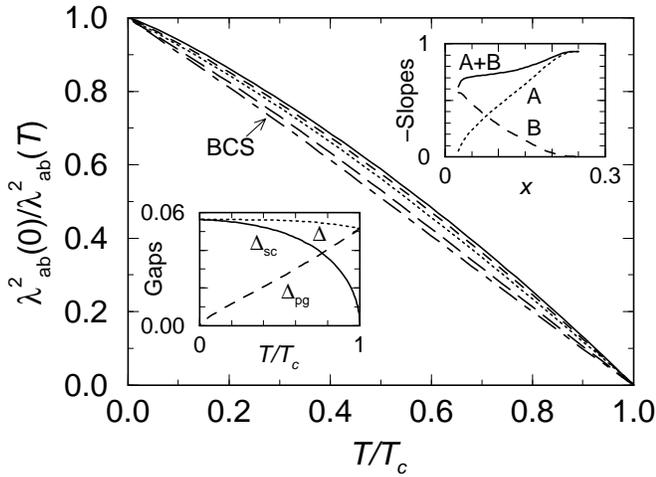}}
\caption{Temperature dependence of the ab-plane inverse squared
  penetration depth.  \textbf{Main figure}: from bottom to top are
  plotted for $x=0.25$ (BCS limit, dot-dashed line), 0.2
  (long-dashed), 0.155 (dotted), 0.125 (dashed) and 0.05 (solid
  line). \textbf{Lower inset}: energy gaps as a function of $T/T_c$ for
  $x=0.125$. \textbf{Upper inset}: (A) the slope given by the low
  temperature expansion assuming a constant $\Delta_{sc}(T)=\Delta
  (0)$; (B) the ratio $\frac{\Delta_{sc}^2(T)}{\Delta^2
  (0)}/\frac{T}{T_c}$ at $T/T_c=0.2$; and (A+B) the sum of two
  contributions.}
\label{Fig2}
\end{figure}

(ii) The superfluid density (normalized to its $T=0$ value), given by
Eq.~(\ref{rhos}), is plotted in Fig.~2 as a function of $T/T_c$ for
several representative values of $x$, ranging from the highly over- to
highly underdoped regimes.  These plots clearly indicate a
``quasi-universal'' behavior with respect to $x$:
$\rho_s(T)/\rho_s(0)$ \textit{vs.}~$T/T_c$ depends only slightly on
$x$. Moreover, the shape of these curves follows closely that of the
weak-coupling BCS theory. The, albeit, small variation with $x$ is
systematic, with the lowest value of $x$ corresponding to the top
curve.
Recent experiments provide some preliminary evidence for this universal
behavior \cite{Hardy,Cambridge}. However, a firm confirmation requires
further experiments on a wider range of hole concentrations, from
extreme under- to overdoped samples \cite{3dxy}.  This universal
behavior appears surprising at first sight\cite{LeeWen,Millis} because
of the strong $x$ dependence in the ratio $T_c/\Delta(0)$ (see Fig.~1).
It should be noted that universality would \textit{not} persist if
$\rho_s(T)/\rho_s(0)$ were plotted in terms of $T/\Delta(0)$.  The
nontrivial origin of this effect has a simple explanation within the
present theory.
At low to intermediate temperatures, it follows from Eq.~(\ref{rhos}) that
$\lambda_{ab}^2 (0)/\lambda_{ab}^2 (T) = \left(\Delta_{sc}^2
(T)/\Delta^2 (0)\right) \left( 1-A T/T_c + {\cal
O}\left[(T/T_c)^2\right]\right) \approx 1-[A+B(T)]\left(T/T_c\right)$,
where terms of order $\left(T/T_c\right)^2$ and higher have been
neglected. Here $A=32 \sqrt{2}\ln{2}
\left(e^2\lambda_{ab}^2(0)/c^2\right)
t_{\parallel}$ $\times \left(T_c/\Delta(0)\right)$ represents the standard
contribution to the linear $T$ dependence of $\rho_s(T)$. The new term
$B(T)=\left(T_c/T\right) \Delta_{pg}^2(T)/\Delta^2(0)$ derives from
the pseudogap contribution and has a weaker than linear $T$ dependence
(as can be inferred from the lower inset in Fig.~3) \cite{SLope-note}. 
For the
purposes of illustration, these two terms are plotted in the upper
inset of Fig.~2 at $T/T_c=0.2$.  Note that the effective (negative)
``slope'' $A+B$ is relatively $x$ independent over the physical range
of hole concentrations.
Physically, the terms $A$ and $B$ are associated with two compensating
contributions, arising from the quasiparticle and pair excitations,
respectively, so that quasi-universal behavior results at low $T$. It
can be shown that the same compensating effect obtains all the way to
$T_c$, as is exhibited in Fig.~2. Thus, the destruction of the
superconducting state comes predominantly from pair excitations at low
$x$, and quasiparticle excitations at high $x$.

\begin{figure}
\centerline{\includegraphics[width=3.4in]{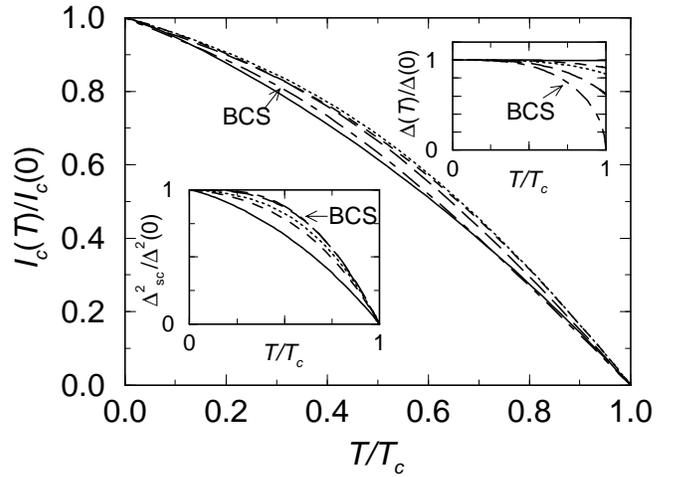}}
\caption{Temperature dependence of c-axis Josephson critical current for
doping concentrations corresponding to the legends 
in  Fig.~\ref{Fig2} (main figure). The variation of $\Delta_{sc}^2$ and
$\Delta$ as a function of $T/T_c$ are plotted in the insets for the
corresponding values of
$x$.
}
\label{Fig3}
\end{figure}

(iii) Finally, as plotted in Fig.~3, we obtain from Eq.~(\ref{crit}),
similarly, unexpected quasi-universal behavior for the normalized
$c$-axis Josephson critical current for the same wide range of $x$ as in
Fig.~2. This behavior is in contrast to the strongly $x$ dependent
quasiparticle tunneling characteristics which can be inferred from the
temperature dependent excitation gap plotted in the upper inset of
Fig.~3.  The origin of this universality is essentially the same as that
for $\rho_s$, deriving from two compensating contributions.
At this time, there do not appear to be detailed studies of $I_c(T)$
as a function of $x$, although future measurements will, ultimately,
be able to determine this quantity. In these future experiments
the quasiparticle tunneling characteristics should be simultaneously
measured, along with $I_c(T)$, so that direct comparison can be made
to the excitation gap; in this way, the predictions indicated in
Fig.~3 and its upper inset can be tested.  Indications, thus
far\cite{JohnZ,Renner}, are that this tunneling
excitation gap coincides rather
well with values obtained from photoemission data (see
Fig.~1).

In summary, in this paper we have proposed a scenario for the
superconducting state of the cuprates. This state evolves continuously
with hole doping $x$, exhibiting unusual features at low $x$ (associated
with a large excitation gap at $T_c$) and manifesting the more
conventional features of BCS theory at high $x$.  In this scenario the
pseudogap state is associated with pair excitations, which act in
concert with the usual quasiparticles. Despite the fact that
the underdoped cuprates exhibit features inconsistent with BCS theory
($T_c /\Delta (0)$ is strongly $x$ dependent and $\Delta$ is finite at
and above $T_c$) we deduce an interesting quasi-universality of the
normalized $\rho_s$ and $I_c$ as a function of $T/T_c$. In these plots
the over- and under-doped systems essentially appear indistinguishable.
Current experiments lend support to this universality in $\rho_s$,
although a wider range of hole concentrations will need to be addressed,
along with future systematic measurements of $I_c$.


We would like to thank D.~A. Bonn, P.~Guptasarma, W.~N. Hardy, M.~Norman
and J.~F. Zasadzinski for useful discussions.  We gratefully acknowledge
the hospitality of the Aspen Center for Physics, where part of this work
was completed.  This research was supported in part by the Science and
Technology Center for Superconductivity funded by the National Science
Foundation under award No. DMR 91-20000.



\begin{references}
\bibitem{PG_fluct}
As expected, $\Delta_{pg}$ is related to the fluctuations of the pair operator,
  $\hat{\Delta}_{\mathbf{q}}\equiv g\sum_{\mathbf{k}}\varphi_{\mathbf{k}}
  c_{-{\mathbf{k}}+{\mathbf{q}}/2\,\downarrow}
  c_{{\mathbf{k}}+{\mathbf{q}}/2\,\uparrow}$, about its mean field value
  $|\Delta_{sc}| = |\langle\hat{\Delta}_{{\mathbf{q}}=0}\rangle|$. Indeed,
  $\langle|\Delta|^2\rangle \equiv \sum_{\mathbf{q}}\langle
  \hat{\Delta}_{\mathbf{q}} \hat{\Delta}^{\dag}_{\mathbf{q}}\rangle =
  g^2\sum_{K,K'}\varphi_{\mathbf{k}}\varphi_{\mathbf{k'}}
  C_2\left(K,K'\right)$, where the irreducible two-particle correlation
  function $C_2\left(K,K'\right) = - \sum_{Q}
  t(Q)\,A(K;Q)\,A\left(K';Q\right)$, with $A(K;Q) =
  G_o(-K+Q/2)\,G(K+Q/2)\,\varphi_{\mathbf{k}}$, in the absence of
  off-diagonal-long-range-order cannot be factorized in the variables $K$ and
  $K'$. This expression for $C_2$ defines the T-matrix $t(Q)$. The combination
  of bare and fully dressed Green's functions enter $A(K;Q)$; for
  $t(Q)=t_{sc}(Q)$, one recovers the Gor'kov factorization
  $C_2\left(K,K'\right)= F(K)\,F^*(K')$, where $F(K)
  =\Delta_{sc}\varphi_{\mathbf{k}}G_o(-K)G(K)$ is the Gor'kov anomalous Green's
  function. Thus, by taking $t(Q)=t_{sc}(Q)+t_{pg}(Q)$, recalling the
  definition of the pair susceptibility $\chi(Q)$, and using the expression for
  the gap equation, one has $\left\langle|\Delta|^2\right\rangle = -\sum_{Q}
  t(Q)\,[g\chi(Q)]^2 = \left|\Delta_{sc}\right|^2 + \left(-\sum_{Q} t_{pg}(Q)
  [g\chi(Q)]^2\right) \approx \left|\Delta_{sc}\right|^2 + \Delta_{pg}^2$,
  where we have exploited once again the fact that, for $T\le T_c$, $t_{pg}(Q)$
  is highly peaked around $Q=0$.
\bibitem{Anisotropy}
We did not fine-tune the ratio $t_\perp/t_\parallel$ since only $T_c$ depends
  slightly on its value (see Ref.~\onlinecite{Qijin}), which is presumably also
  doping dependent.
\bibitem{3dxy}
The experimentally observed deviation from a BCS-like, linear $T$ dependence
  near $T_c$ (see Ref.~\onlinecite{Hardy}) has been associated with 3D~XY
  fluctuations, which are beyond the scope of the present theoretical approach.
\bibitem{SLope-note}
It can be shown that at low temperatures $\Delta^2_{pg}(T)/\Delta^2(0) \propto
  \left(T/T_c\right)^{3/2}$, hence $B(T)$ has a weak $T$ dependence ($\propto
  \sqrt{T/T_c}$\,). This term becomes less important for $T<0.05T_c$, although
  in this regime impurity effects will obscure the linear $T$ dependence.
\end{references}
\end{document}